# METHOD FOR ANALYZING THE SPATIAL DISTRIBUTION OF GALAXIES ON GIGAPARSEC SCALES. II. APPLICATION TO A GRID OF THE HUDF-FDF-COSMOS-HDF SURVEYS

*N.V.NABOKOV and Yu.V.BARYSHEV [1]*

*Using the deep fields of COSMOS, FDF, HUDF, and HDF-N as an example, we discuss the prospects for and limitations on the method for searching for super large structures in the spatial distribution of galaxies proposed in the preceding article of this series. An analysis of the distribution N(z) of photometric redshifts in a grid of the deep fields of HUDF-FDF-COSMOS-HDFN reveals the possible existence of super large structures with a contrast dN/N~50% and tangential and radial dimensions of about 1000 Mpc. The reality of the detected candidate super large structures in the universe can be verified by further observations with a finer grid of deep fields. The influence of systematic errors can be reduced by observing the same deep fields with several 3-10 meter telescopes and utilizing different methods for determining the photometric redshifts.*

## 1. Introduction

In the first article of this series [1] we proposed a method for determining the dimensions and contrast of super large inhomogeneities in the large scale structure of the universe over a wide range of redshifts. Using this method on deep field observations can set direct observational limits on the existence of super large structures in the spatial distribution of galaxies for redshifts of 0.5 - 5.

As a first step in applying this method to the search for super large structures [1], in this article we examine a grid of fields formed by the deep COSMOS, FDF, HUDF, and HDF-N surveys. Section 2 is an analysis of the radial distribution of the galaxies in the deep fields of COSMOS, FDF, HUDF, and HDF-N. The possibility of detecting regions with enhanced or reduced densities of visible matter on scale lengths of up to 1000 Mpc is demonstrated there. In section 3 the possible tangential dimensions of super large structures are estimated on the basis of the angular separations between individual deep fields. Additional tests of the reality of super large structures in the universe are discussed in section 4. The basic results are summarized in the concluding section 5.

---

[1] Institute of Astronomy , Saint Petersburg State University, Russia; e-mail: NabokovNikita@yandex.ru; yubaryshev@mail.ru

## 2. Radial distributions of galaxies in the deep fields of COSMOS, FDF, HUDF, and HDF-N

### 2.1. Parameters of the fields of COSMOS, FDF, HUDF, and HDF-N.

As an example of applying the method of three dimensional space tomography [1] we shall examine the distribution of photoelectric redshifts in the COSMOS (Cosmic Evolution Survey [2]), FDF (FORS Deep Field of the ESO VLT [3]), HUDF (Hubble Ultra Deep Field [4]), and HDF-N (Hubble Deep Field North [5]) deep galactic surveys. The observed distributions of galactic redshifts in bins of size $\Delta z = 0.1, 0.2$, and $0.3$ for these fields are compared with the expectations for a uniform spatial distribution in a sample that is limited in terms of visible stellar magnitude. Table 1 lists some parameters of these fields.

TABLE 1. Basic Parameters of the COSMOS, FDF, HUDF, and HDF-N Fields Studied in this Paper

| Field | $\alpha$ | $\delta$ | Angular size | $m_{\lim}$ (filter B) |
|---|---|---|---|---|
| COSMOS | $10^h 00^m$ | $+02°12'$ | $77' \times 77'$ | 25 |
| FDF | $01^h 06^m$ | $-25°46'$ | $7' \times 7'$ | 27 |
| HUDF | $03^h 32^m$ | $-27°47'$ | $3' \times 3'$ | 30 |
| HDF-N | $12^h 36^m$ | $+62°13'$ | $2.3' \times 2.3'$ | 29 |

Tables 2-5 show the observed candidate super large structures, together with the values of the expected mean square deviations for Poisson fluctuations and for correlated structures with parameters $r0 = 5$ Mpc and $\gamma = 1.8$ in the cases of the COSMOS, FDF, HUDF, and HDF-N fields for bins $\Delta z = 0.3$ according to Eqs. (6,I)-(10,I). (Here and in the following the equations refer to those in part I of this paper [1].) The basic problem, which remains to be solved, is to separate the fluctuations associated with structures from the fluctuations owing to systematic errors in the method for determining photo-z.

### 2.2 N(z) for the COSMOS sample.

At present the COSMOS survey is the largest deep multiband survey of galaxies. It contains about half a million photometric redshifts evaluated using 30 filters for 607617 galaxies with i < 26 [6]. Since April 2008, a catalog of the photometric redshifts for 385065 galaxies in a sample limited in visible stellar magnitude to i < 25 has been available on the internet [7].

The accuracies of the estimates of photo-z for z < 1.25 are [6] $\sigma_z = 0.02, 0.04, 0.07$ for i ~ 24, i ~ 25, and i ~ 25.5, where the Balmer limit (at 4000 E) does not go beyond about 9000 E. For z > 1.25 the accuracy falls to $\sigma_z \approx 0.14$ (i ~ 24) and reaches $\sigma_z \approx 0.19$ for z ~ 2.2.

For z > 2.5 the accuracy improves to $\sigma_z \leq 0.1$ (24 < i < 25), where the Balmer limit shifts into the J filter (z ~ 2). Figure 1 shows the radial distribution N(z) for a bin size of 0.3. The observed deviations from the theoretical radial distribution (5, I) are shown in Fig. 2 for bin widths $\Delta z = 0.1, 0.2,$ and $0.3$.

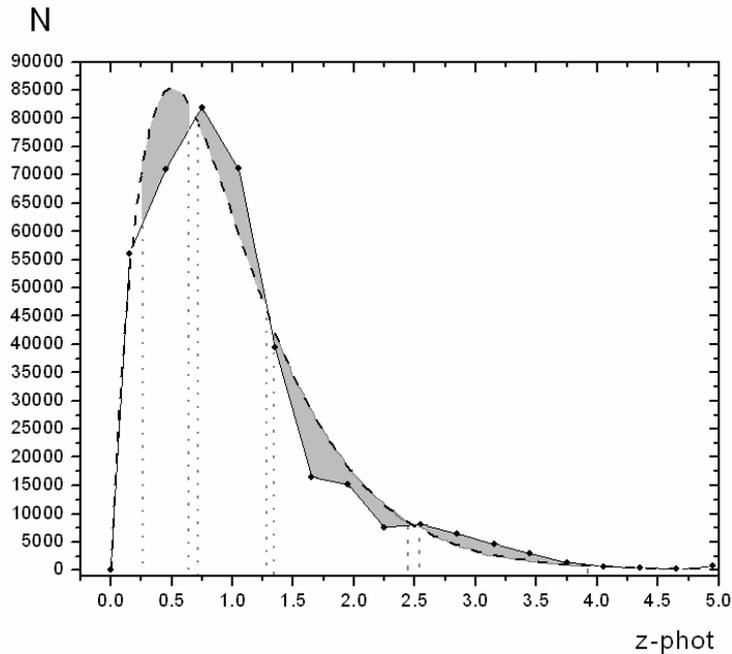

Fig. 1. The radial redshift distribution of galaxies in the COSMOS survey and the possible super large regions with elevated and reduced density for $\Delta z = 0.3$. The dashed curve is the expected redshift distribution for a uniform spatial distribution of the galaxies. The number of galaxies is 382143.

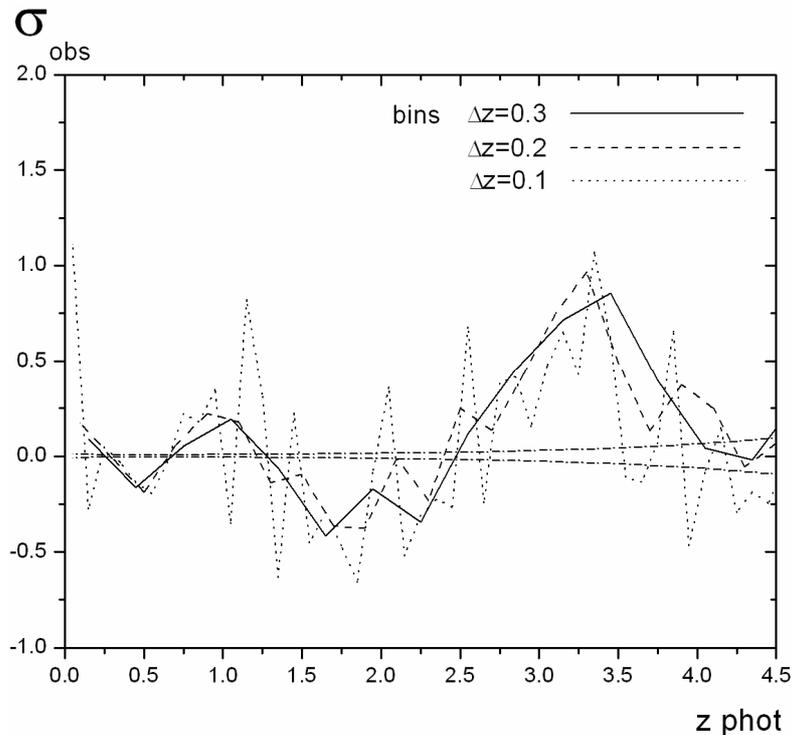

Fig. 2. The observed deviations from a uniform distribution for the radial redshift distributions of galaxies in the COSMOS survey for bin sizes $\Delta z = 0.1, 0.2,$ and $0.3$ (dotted, dashed, and smooth

curves, respectively). The dot-dashed curve is the Poisson noise $\sigma_p$ for $\Delta z = 0.3$. The number of galaxies is 382143.

Table 2 lists the candidate regions of enhanced and reduced galactic concentration in the deep field of COSMOS. According to Fig. 2, the relative deviations $\sigma_{obs}$ are up to 50%, and their sizes may be a large as 1000 Mpc.

TABLE 2. Structures Resolved in a Radial Distribution for the COSMOS Survey

| Designation | $z_{start}$ | $z_{finish}$ | $\sigma_P$ | $\sigma_{corr}$ | Size (Mpc) | Contrast (modulus) |
|---|---|---|---|---|---|---|
| COSMOS-SLV-1 | 0.27 | 0.64 | 0.003 | 0.101 | 1203 | 0.16 |
| COSMOS-SLC-1 | 0.7 | 1.27 | 0.004 | 0.063 | 1359 | 0.18 |
| COSMOS-SLV-2 | 1.32 | 2.44 | 0.007 | 0.048 | 1683 | 0.42 |
| COSMOS-SLC-2 | 2.5 | 3.99 | 0.019 | 0.041 | 1291 | 0.82 |

**2.3. N(z) for the FDF sample.**

About 7000 photometric redshifts of galaxies were measured in the deep field of FDF [3] using UBgRIJK filters. We have used these data to construct the distribution of redshifts and the number of deviations for bin sizes of $\Delta z = 0.1, 0.2$, and $0.3$. Figure 3 shows the parameters of the radial distribution (5, I): $A = 2455.67$, $\alpha = 0.82$, $\beta = 1.14$, and $z_0 = 1.03$ for a $\Delta z = 0.3$ bin.

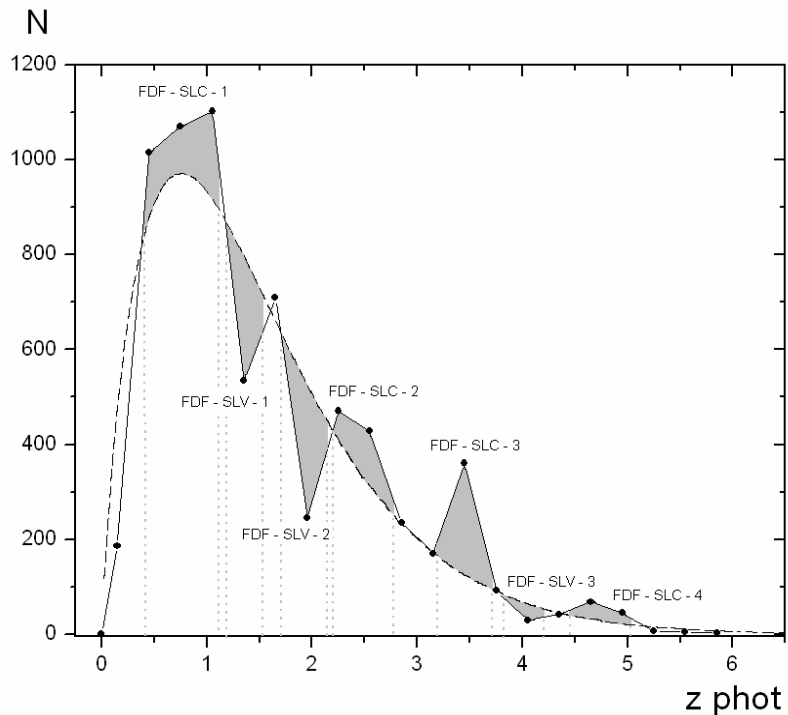

Fig. 3. The radial redshift distribution of galaxies in the FDF field and the possible super large regions with elevated and reduced density for $\Delta z = 0.3$. The dashed curve is the expected redshift distribution for a uniform spatial distribution of the galaxies. The number of galaxies is 6815.

Table 3 lists the candidate regions of enhanced and reduced galactic concentration in the deep field of FDF. The observed relative deviation (10, I) of the number of galaxies from the expected uniform distribution in a sample of galaxies limited in terms of visible stellar magnitude for the FDF field is shown in Fig. 6 (see section 3) for a bin width of $\Delta z = 0.3$.

TABLE 3. Structures Resolved in the Radial Distribution for the FDF Field

| Designation | $z_{start}$ | $z_{finish}$ | $\sigma_P$ | $\sigma_{corr}$ | Size (Mpc) | Contrast (modulus) |
|---|---|---|---|---|---|---|
| FDF-SLC-1 | 0.4 | 1.12 | 0,033 | 0,548 | 1966 | 0.20 |
| FDF-SLV-1 | 1.18 | 1.56 | 0,044 | 0,416 | 728 | 0.33 |
| FDF-SLV-2 | 1.71 | 2.18 | 0,057 | 0,378 | 675 | 0.53 |
| FDF-SLC-2 | 2.21 | 2.8 | 0,090 | 0,344 | 664 | 0.37 |
| FDF-SLC-3 | 3.16 | 3.73 | 0,125 | 0,329 | 455 | 1.88 |
| FDF-SLV-3 | 3.83 | 4.3 | 0,178 | 0,317 | 309 | 0.53 |
| FDF-SLC-4 | 4.4 | 5.09 | 0,033 | 0,548 | 377 | 1.14 |

**2.4. N(z) for samples from HUDF.**

The superdeep Hubble field HUDF has been studied using BVizJH photometry [8]. The J and H filters of the NICMOS infrared camera were used to determine the photometric redshifts. Using their own programs, the authors of Ref. 8 determined the photo-z for 7560 galaxies.

In constructing a model distribution N(z) we selected galaxies with a likelihood of > 70% for the determination of photo-z, so that a sample of 5446 galaxies with photoelectric redshifts was obtained. Figure 4 shows the distribution of the photo-z galaxies in HUDF [8]. The parameters of the uniform model radial distribution are A = 464396.13, α = 0.71, β = 1.03, and $z_0$ = 1.86. Table 4 lists the candidate regions of enhanced and reduced galactic concentrations in the deep field of HUDF. Figure 6 shows the relative deviations from the expected for the uniform distribution.

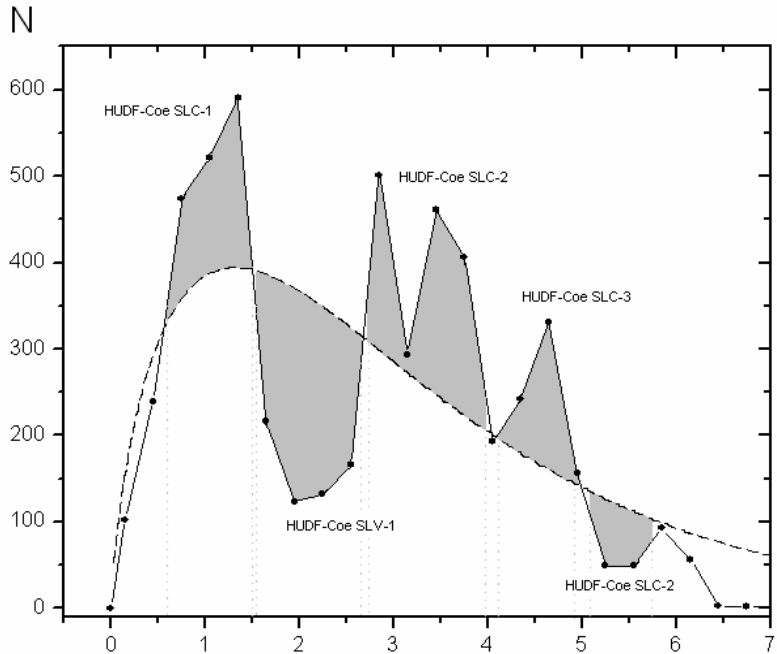

Fig. 4. The radial redshift distribution of 5446 galaxies in the HUDF field for the sample of Ref. 8 and the possible super large regions with elevated and reduced density for $\Delta z = 0.3$. The dashed curve is the expected redshift distribution for a uniform spatial distribution of the galaxies. The number of galaxies is 5446.

TABLE 4. Structures Resolved in the Radial Distribution for the HUDF Field

| Designation | $z_{start}$ | $z_{finish}$ | $\sigma_P$ | $\sigma_{corr}$ | Size (Mpc) | Contrast (modulus) |
|---|---|---|---|---|---|---|
| HUDFcoe-SLC-1 | 0.61 | 1.48 | 0.051 | 0.451 | 2019 | 0.42 |
| HUDFcoe-SLV-1 | 1.54 | 2.65 | 0.052 | 0.342 | 1507 | 0.62 |
| HUDFcoe-SLC-2 | 2.7 | 3.13 | 0.056 | 0.305 | 412 | 0.61 |
| HUDFcoe-SLC-3 | 3.17 | 3.99 | 0.061 | 0.290 | 628 | 0.78 |
| HUDFcoe-SLC-3 | 4.18 | 4.95 | 0.079 | 0.261 | 442 | 0.98 |

**2.5. N(z) for the HDF-N sample.**

Photometric data in ubvr filters are available on the internet [5]. Here we processed these data from selecting objects (SExtractor) to evaluating photo-z (Hyper z). The initial catalog of observed galaxies contains 3301 galaxies. On applying the procedure for determining photo-z and imposing the condition of a likelihood >70%, we obtained a catalog of 1916 galaxies. For determining photo-z we used 4 band photometry (filters with effective wavelengths (E)): u ($\lambda_{u,eff} = 3011.6$), b ($\lambda_{b,eff} = 4573.6$), v ($\lambda_{v,eff} = 6033.9$), and r ($\lambda_{r,eff} = 8009.1$). The theoretical radial distribution satisfies Eq. (5.I), where the parameters were found by the method of least squares.

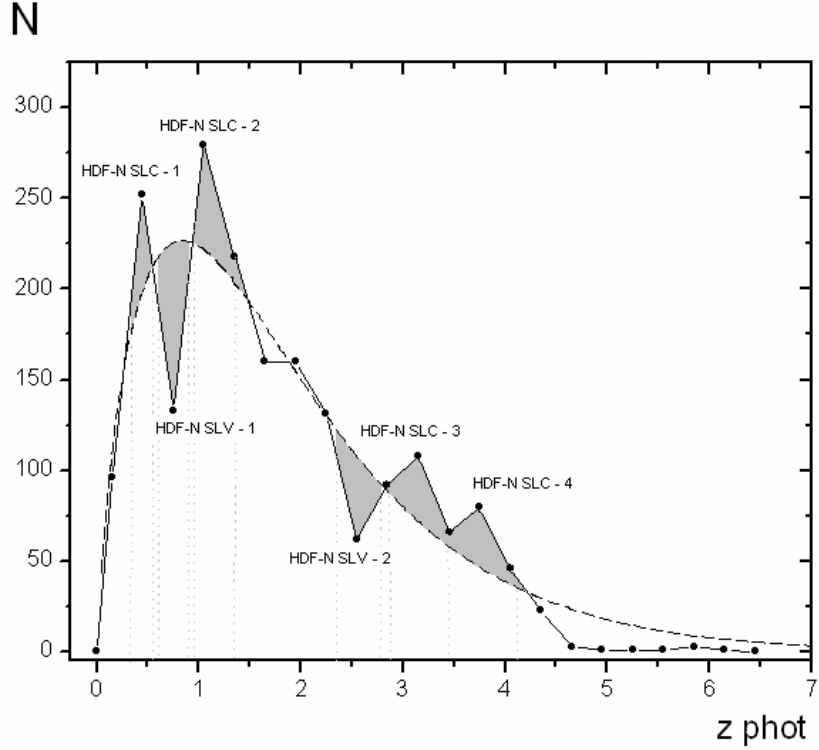

Fig. 5. The radial redshift distribution of galaxies in the HDF-N field and the super large regions with elevated and reduced density for Δz = 0.3. The dashed curve is the expected redshift distribution for a uniform spatial distribution of the galaxies; the smooth curve is the observed distribution. The number of galaxies is 1916.

TABLE 5. Structures Resolved in the Radial Distribution for the HDF-N Field

| Designation | $z_{start}$ | $z_{finish}$ | $\sigma_P$ | $\sigma_{corr}$ | Size (Mpc) | Contrast (modulus) |
|---|---|---|---|---|---|---|
| HDF-N-SLC-1 | 0.3 | 0.54 | 0.071 | 0.886 | 796 | 0.27 |
| HDF-N-SLV-1 | 0.59 | 0.9 | 0.069 | 0.771 | 848 | 0.41 |
| HDF-N-SLC-2 | 0.95 | 1.35 | 0.069 | 0.517 | 865 | 0.26 |
| HDF-N-SLV-2 | 2.32 | 2.76 | 0.088 | 0.394 | 485 | 0.43 |
| HDF-N-SLC-3 | 2.89 | 3.42 | 0.106 | 0.364 | 466 | 0.5 |
| HDF-N-SLC-4 | 3.42 | 4.09 | 0.131 | 0.344 | 484 | 0.72 |

Figure 5 shows the observed and theoretical radial distributions for the HDF-N field in redshift bins Δz = 0.3.

The dotted curve shows the theoretical distribution for parameters A = 584.68, α = 0.81, β = 1.01, and $z_0$ = 1.04 (according Eq. (5, I)) and the smooth curve, the observed distribution. Table 5 lists the candidate regions with enhanced and reduced galactic densities in the deep field of HDF-N. Figure 7 shows the relative deviations from the expected of a uniform distribution.

## 3. Comparison of the radial distributions of galaxies in a grid of deep fields

The final step in the proposed method [1] of searching for super large structures is to compare the radial distributions in neighboring directions of the sky corresponding to the existing grid of deep fields.

When the sizes of the structures exceed the angular distance between the corresponding deep fields, the deviations from uniformity must be examined with slight shifts along the z axis. In this way it is possible to obtain an estimate for the tangential linear dimensions of super large structures in the spatial distribution of galaxies on scales of thousands of Mpc. In our case we have 4 fields, COSMOS, FDF, HUDF, and HDF-N. Thus, for example, the fields of HUDF and FDF lie at an angular distance of 36° from one another, which corresponds to roughly 1700 Mpc/h for z = 1.

In Figs. 6 and 7 the relative deviations $\sigma_{obs}(z)$ for the COSMOS, FDF, HUDF, and HDF-N fields are compared with a redshift step size of 0.3. Tables 2-5 list the observed candidate super large structures — possible regions with elevated and reduced concentrations of galaxies relative to the Poisson noise $\sigma_p$.

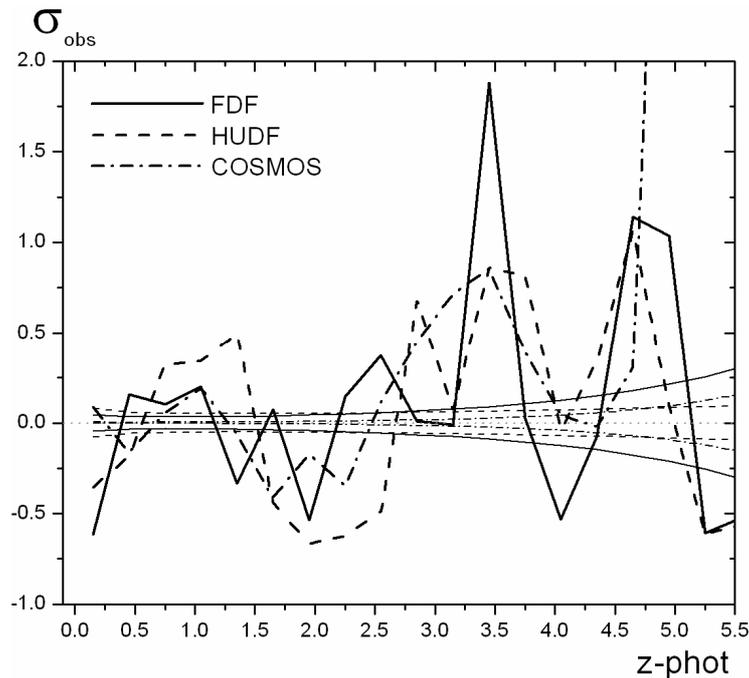

Fig. 6. The observed deviations and Poisson noise for the FDF, HUDF, and COSMOS fields. The thick lines indicate the observed deviations and the thin lines, the Poisson noise $\sigma_p$ or the FDF, HUDF, and COSMOS fields.

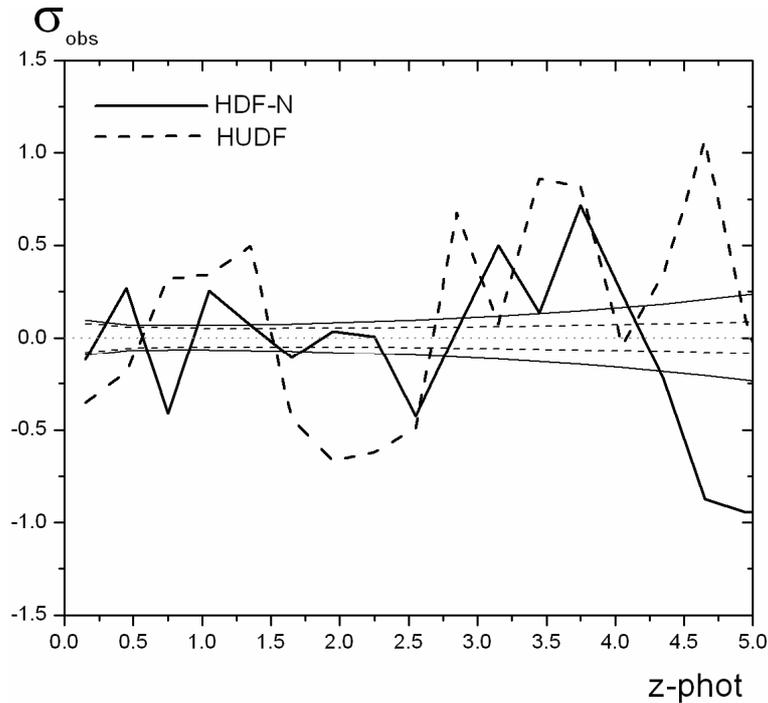

Fig. 7. The observed deviations and Poisson noise for HDF-N and HUDF. The thick dashed trace is the observed deviations and the thin dashed trace is the Poisson noise $\sigma_p$ or the HUDF field; the thick smooth trace is the observed deviations and the thin smooth trace is the Poisson noise $\sigma_p$ for the HDF-N field.

Figure 6 shows that there is a similarity in the behavior of the observed fluctuations on super large scales in the fields of COSMOS, HUDF, and FDF. Since the COSMOS, HUDF, and FDF fields were observed with different instruments and have been processed by different techniques, it seems possible that the contribution to the resultant photo-z from real in homogeneities may exceed that from selection effects and systematic errors. Substructures with large contrasts in density show up in the galactic redshifts distributions constructed with minimal bin widths in all the fields.

The radial distribution N(z) and the deviations from uniformity found for the COSMOS survey are the most reliable; there the number of galaxies is hundreds of thousands and the accuracy of the photo- z measurements is better than 0.1 for z < 4.5. Thus, the similarity of the regions of elevated and reduced density in the fields of COSMOS-HUDF-FDF can be interpreted as a consequence of super large structures extending by up to 3000 Mpc in a transverse direction. The observed shifts of the structures along z in the different fields may correspond to a real radial shift in the super large structures as the direction of the line of sight is changed.

On the other hand, a comparison of the radial distributions of the HDF-N and HUDF fields shows that no correlations in the structures show up within the z interval 1.5-2.5. (See Fig. 7.) This may be an indication of a limited transverse size for a super large structure that is visible in the COSMOS-HUDF-FDF fields.

## 4. Observational tests of the reality of super large structures

In accordance with Eq. (11, I), the observed deviations from a uniform distribution in the COSMOS, FDF, HUDF, and HDF-N deep fields contain contributions from Poisson fluctuations, correlated structures, and systematic errors. A necessary condition for the reliable detection of super large structures is both the Poisson and systematic errors be small. The above analysis shows that the Poisson errors are smaller than the observed deviations, but the magnitude of the systematic errors is still an open question and requires further study. Since the problem of quantitatively estimating the systematic errors is still unsolved, additional arguments supporting the reality of super large structures play an important role.

**4.1. Real clusters of galaxies in spectral observations.**

Important observational evidence of the reality of large fluctuations in the distribution of galaxies with respect to their photometric redshifts is provided by spectral observations of galaxies in the fields of COSMOS [9,10] and FDF [3].

According to Fig. 2 of Ref. 9 and Fig. 5 of Ref. 10, radial inhomogeneities with dimensions of about 1000 Mpc can be seen in the distribution N(z) for a sample of 104 galaxies with spectrally measured redshifts; they are interpreted as "large cosmic variance".

According to Ref. 3, the distribution of measured spectral redshifts of 340 galaxies in the FDF field reproduces the distribution of photometric redshifts, including noticeable maxima in the galactic density at redshifts of 0.3, 0.8, 2.4, and 3.4, which, the authors of that article believe is an obvious reflection of the reality of large structures "like a sponge" in the spatial distribution of the galaxies. Our analysis of the distribution of photo-z in the FDF field based on the criterion $|\Delta N / N| > +\sigma_P$ also reveals the presence of super large structures with peaks at z around 2.5, 3.0, and 4.7. In our case, however, these are local maxima inside of enormous structures extending over z within the ranges of 2.1-3.7 and 4.2-5.2.

**4.2. Other deep surveys.**

The ALHAMBRA project [11], in which observations of 8 deep fields in 20 filters are planned with an overall number of $6.6 \times 10^5$ galaxies, is currently already in the process of completion.

In order to study systematic effects, artificial catalogs of uniformly distributed galaxies with parameters corresponding to those of the surveys must be compiled to accompany the deep field observations, so that selection effects and systematic distortions in the observed photometric redshift distributions can be evaluated quantitatively.

Observations of the redshift distributions of gamma bursts can also serve as a grid that covers the entire sky. Available data on redshift measurements for more than 100 SWIFT gamma bursts are compatible with the existence of super large structures in the distribution of the parent galaxies of the GRB [12]. Field grids can also be assigned to known deep fields with centers in the parent galaxies of gamma bursts [13-15].

## 5. Conclusion

In recent years there has been a noticeable trend in observational cosmology toward the discovery of structures of every larger size in the spatial distribution of galaxies. In this paper we have used the method proposed in Ref. 1 to show that the photometric redshift distributions of galaxies in a grid of deep samples of galaxies can be used as an instrument for studying super large structures with sizes of thousands of Mpc in the universe.

An analysis of the photometric redshift distributions of galaxies in the deep fields of COSMOS, FDF, HUDF, and HDF-N has shown that the observed fluctuations in the number of galaxies in large redshift bins (0.1-0.3) are considerably greater than the $\sigma_p$ level, so that they may be caused by correlated structures. Systematic effects, which require further study, may also make a significant contribution to the observed fluctuations.

It is interesting that the relative density fluctuations in the COSMOS-HUDF-FDF fields do behave similarly. Since the data for these fields were obtained with different instruments and processed using different programming systems, it seems possible that real super large structures are making a significant contribution to the observed fluctuations. Thus, we presume that the observed fluctuations may also contain a signal corresponding to super large structures in the spatial distribution of the galaxies with scale lengths on the order of gigaparsecs. For example, the HUDF and FDF fields are separated from one another in the sky by 36 degrees, so that the transverse size of a super large structure at a distance of z = 1 would be about 1700 Mpc/h.

The existence of super large structures is also consistent with the already known large structures in the universe discovered by different observational techniques. For example, a structure of size roughly 500 Mpc/h that was discovered in the SDSS survey (the Sloan Great Wall [16]) is well known. Based on redshift catalogs accessible up to 1998, evidence has been found [17] for the existence of structures with sizes of up to 1000 Mpc/h. An analysis [18] of the SDSS LRG DR5 catalog revealed large fluctuations in the concentration of galaxies on scales of 100-300 Mpc/h. Photometric redshifts in the SDSS LRG galactic sample have been used [19] to obtain an exponential dependence for the power spectrum extending to scale lengths λ = 1200 Mpc.

We note that the unexpected discovery of a large scale "dark flow," based both on observations of the Syunyaev-Zeldovich effect for x-ray galactic clusters [20] and on observations of the peculiar velocities of the galaxies [21], means that the entire local volume, of size 300 Mpc/h, is undergoing a large scale motion, a fact consistent with the existence of super large structures on scales of 1000 Mpc/h.

Future research dealing with the spatial distribution of galaxies on gigaparsec scales will require

- organizing observations of a grid of deep multiband surveys covering a large portion of the celestial sphere with a cell size of ~10°×10° with fields of size ~10′×10′ at their nodes;
- the use of different telescopes and methods of evaluating the photo-z for identical deep fields;
- the use of model radial distributions of galaxies obtained with LCDM models for the evolution of large scale structure for redshifts in the range 0.1 - 6;
- studies of the redshift distribution of gamma bursts (GRB) in different regions of the sky; and,
- the compiling of artificial catalogs of uniformly distributed galaxies in which the process of observing the deep fields and the systematic effects associated with evaluating the photometric redshifts are modeled.

We thank a reviewer for useful comments that have greatly improved the presentation. We also acknowledge partial financial support from the Foundation for Leading Scientific Schools, grant No. NSh 1318.2008.2, and the Russian Foundation for Basic Research (RFFI), grant No. 09-02-00-143.